\documentclass[journal=nalefd,manuscript=letter]{achemso}

\usepackage{graphicx}
\usepackage{bm}
\usepackage{amsmath}
\usepackage{amsfonts}
\usepackage{amssymb}
\usepackage{braket}
\usepackage{mathrsfs}
\usepackage{color}
\usepackage[colorlinks,linkcolor={black},citecolor={black},urlcolor={black}]{hyperref}

\newcommand{\bs}{\boldsymbol}
\newcommand{\mc}{\mathcal}

\author{Huiyuan Zheng}
\affiliation[University of Hong Kong]{New Cornerstone Science Laboratory, Department of Physics, The University of Hong Kong, Hong Kong, China}
\alsoaffiliation{HKU-UCAS Joint Institute of Theoretical and Computational Physics at Hong Kong, Hong Kong, China}
\altaffiliation{These authors contributed equally to this work.}

\author{Dawei Zhai}
\affiliation[University of Hong Kong]{New Cornerstone Science Laboratory, Department of Physics, The University of Hong Kong, Hong Kong, China}
\alsoaffiliation{HKU-UCAS Joint Institute of Theoretical and Computational Physics at Hong Kong, Hong Kong, China}
\altaffiliation{These authors contributed equally to this work.}

\author{Cong Xiao}
\affiliation{HKU-UCAS Joint Institute of Theoretical and Computational Physics at Hong Kong, Hong Kong, China}
\alsoaffiliation{Institute of Applied Physics and Materials Engineering, University of Macau, Taipa, Macau, China}
\email{congxiao@um.edu.mo}

\author{Wang Yao}
\affiliation[University of Hong Kong]{New Cornerstone Science Laboratory, Department of Physics, The University of Hong Kong, Hong Kong, China}
\alsoaffiliation{HKU-UCAS Joint Institute of Theoretical and Computational Physics at Hong Kong, Hong Kong, China}
\email{wangyao@hku.hk}

\title
  {Interlayer electric multipoles induced by in-plane field from quantum geometric origins}


\begin{document}

\begin{tocentry}
\centering
\includegraphics[width=0.7\textwidth]{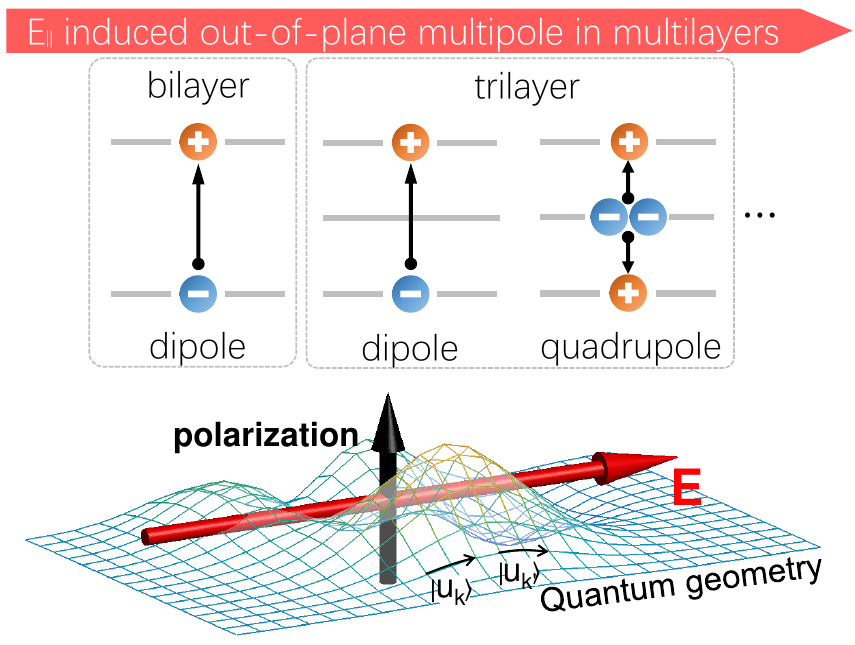}
\end{tocentry}

\begin{abstract}
We show that interlayer charge transfer in 2D materials can be driven by an in-plane electric field, giving rise to electrical
multipole generation in linear and second order of in-plane field. The linear and nonlinear
effects have quantum geometric origins in the Berry curvature and quantum metric respectively, defined in extended parameter spaces characteristic of layered materials. We elucidate their symmetry characters, and demonstrate sizable dipole and quadrupole polarizations respectively in
twisted bilayers and trilayers of transition metal dichalcogenides. Furthermore, we show that the effect is strongly enhanced during the topological phase transition tuned by interlayer translation. The
effects point to a new electric control on layer quantum degree of freedom.

\end{abstract}


The advent of 2D materials has provided opportunities for pseudospintronics, the electronics concept aiming at utilizing electron's
emergent quantum degrees of freedom, such as {\it valley} in the momentum space and {\it layer} in van der Waals (vdW)
structures~\cite{Pesin2012,Xu2014,banerjee2009bilayer,Prada2009,Gong2013}.
Compared to spin and valley, the layer pseudospin features uniquely an electrical dipole that allows its direct coupling to out-of-plane electric field. Layer polarization can therefore be generated by an interlayer bias, including in metallic few-layers~\cite{Fei2018}. 
Besides such control based on electrostatics~\cite{Young2011}, however,
very limited approaches are known for addressing the layer polarization~\cite{stern2021interfacial,woods2021charge,yasuda2021stacking}, in sharp contrast to the versatile possibilities for valley and spin in transition metal dichalcogenides (TMDs) and graphene by exploiting the associated quantum geometric
properties~\cite{Xu2016}.
Notably, intriguing layer pseudospin textures can emerge in moir\'e superlattices and thereby introducing quantum geometric properties of electrons~\cite{wu2019topological,yu2019giant,ZhaiPRM2020,devakul2021magic,Gao2020,zhai2023time,Chen2023}, parallel to the conventional approach exploiting spin-orbit coupling~\cite{Manchon2019}. 

The layer pseudospin also expands the parameter space of Bloch states, enabling exploring quantum geometry beyond the $k$ space.
A novel band geometric quantity characteristic of vdW layered structures was revealed in Ref.~\cite{Chen2023}, which contributes an intrinsic nonlinear Hall effect in nonmagnetic chiral bilayers. This contrasts with the recently observed intrinsic nonlinear Hall effect in antiferromagnets, which is attributed to the $k$-space Fubini-Study metric~\cite{wang2023quantum,gao2023quantum}. Despite such progress, the study of quantum geometric effects in the extended space unique to layered structures is still in its infancy, more geometric quantities and their nontrivial consequences yet to be explored.

\begin{figure}[t]
	\includegraphics[width=0.7\textwidth]{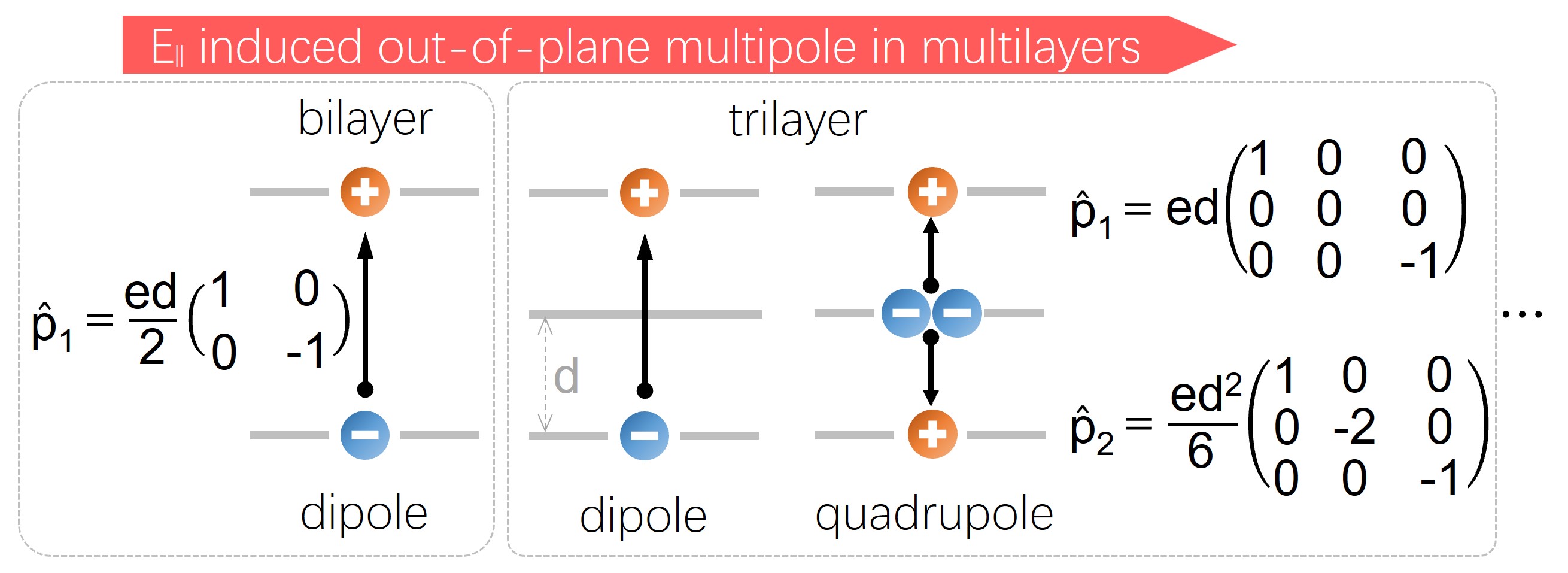} \caption{Schematics of
		few-layers and interlayer multipoles. 
		An in-plane electric field can generate these charge multipoles. Dipole ($\hat{p}_1$) and quadrupole ($\hat{p}_2$) moment operators are given in the layer space for bilayer and trilayer.}%
	\label{fig_multipole}%
\end{figure}

Here we unveil a band geometric effect through which an in-plane electric field can efficiently drive layer polarization in vdW layered structures. This is reminiscent of electrically-driven spin polarization~\cite{Manchon2019}, whereas `high-spin' configurations can be readily achieved here, corresponding to layer number, which allows exploration of electric quadrupole and higher multipoles (Fig.~\ref{fig_multipole}). Such layer polarization responses are \textit{anomalous}, which can be expressed in terms of intrinsic band geometric quantities defined in extended parameter spaces characteristic of layer hybridized electrons. Specifically, the linear and nonlinear responses are connected to Berry curvature and Berry connection polarizability (BCP) dipoles, respectively. Interestingly, the latter is closely connected to the quantum metric~\cite{provost1980riemannian}. Therefore, the linear and nonlinear interlayer multipole generations make it possible to explore both the phase and distance aspects of the quantum geometry. We also clarify the symmetry characters of responses and show that they can be tailored by controlling the symmetries of the vdW materials.

As specific examples, we show sizable interlayer electric dipole generation in twisted TMD bilayer with uniaxial strain. Generalized to trilayers, rich linear and nonlinear responses of dipole and quadrupole emerge under the control of interlayer translation, and the size of the effects is shown to be comparable or even larger than that driven by optical means~\cite{Gao2020,XiaoRC2022,Song2023}.
Our work not only offers a new method of controlling layer polarization in vdW layered materials, the various interlayer multipole responses can also serve as experimental probes of novel band geometric quantities unique to layered materials.

\emph{Anomalous layer polarization of layered electronic states.-}
Central to the proposed effect is the layer polarization of
individual electrons under an in-plane field $\bs{E}$. We
proceed by the semiclassical theory~\cite{Xiao2010,Gao2014,Gao2019,Xiao2021Adiabatic,Xiao2022MBT}, which
has successfully described various response phenomena dictated by band geometric
properties. One prominent example is the interpretation of the anomalous Hall effect via the \textit{anomalous velocity} acquired by an electron under $\bs{E}$ field~\cite{Gao2014}. 
Here, there is also an \textit{anomalous layer polarization} induced by $\bs{E}$ field for electrons residing on coupled layers, akin to the consequence of anomalous charge motions in the out-of-plane direction.

Electric field induced layer polarization can be visualized as the generation of charge-neutral out-of-plane electric multipoles that conserves the total charge (Fig.~\ref{fig_multipole}), which is a characteristic of layertronics that does not have counterparts in spintronics. In the following we will use the phrases layer polarization and multipole generation interchangeably. 
Notice that such interlayer multipoles in quasi-2D layered materials behave like pseudospins due to a finite and discrete number of layers, in contrast to their counterparts in 3D crystals~\cite{MultipoleScience2017,MultipolePRB2017}. 
In this work we will focus on bilayers and trilayers, hence dipole and quadrupole generations, but the discussions can be extended to higher multilayers and multipoles.
The forms of the interlayer multipoles, and their operators $\hat{p}_{i}$ in the layer space (Fig.~\ref{fig_multipole}), are rooted in the multipole expansion of the electrostatic energy in a layered structure (see Supporting Information).

Formally, the amount of interlayer multipoles in a crystal can be evaluated by counting the contributions from Bloch states, $p_i(n,\bs{k}) = \braket{u_{n\bs{k}}|\hat{p}_{i}|u_{n\bs{k}}}$, where $\ket{u_{n\bs{k}}}$ denotes the periodic part of the Bloch state with band index $n$ and crystal momentum $\boldsymbol{k}$. Multipole generation is expected after applying an $\bs{E}$ field, due to its correction of $\ket{u_{n\bs{k}}}$ that changes the charge occupation among different layers.
A convenient approach for evaluating the responses of a generic observable can be introduced by considering the coupling term between the observable and its corresponding field~\cite{Dong2020}.
In the current case of interlayer multipole generation, the coupling term is $H_{c}=-\sum_{i=1,2,\cdots}\hat{p}_{i}\lambda_{i}$, where $\lambda_{i}$ denotes the coupling field to multipole $\hat{p}_{i}$. Specifically, a dipole ($\hat{p}_1$) couples to an out-of-plane electric field ($\lambda_{1}$), a quadrupole ($\hat{p}_2$) couples to the gradient of the out-of-plane electric field ($\lambda_{2}$), etc.
From the action of a wave packet and its variation with respect to $\lambda_{i}$, the response coefficients of $\hat{p}_i$ can be obtained by taking $\lambda_{i}\rightarrow0$ in the final expressions~\cite{Dong2020,Gao2020,Xiao2022MBT}. All the results can be expressed in terms of the intrinsic quantities of the system under study. This approach is advantageous in revealing the geometric origin of the responses.

According to such a semiclassical theory accurate to the second order of $\bs{E}$ field, the layer polarization contributed by individual electrons is given by (details in Supporting Information)
\begin{equation}
\begin{aligned}
p_{i}\left(  \boldsymbol{k}\right)
&=p_{i}^{0}(\bs{k})+p_{i}^{\text{L}}(\bs{k})+p_{i}^{\text{NL}}(\bs{k})\\
&=-\partial_{\lambda_{i}}\left(
\varepsilon+\delta\varepsilon\right)  +(\boldsymbol{\Omega}_{\lambda
_{i}\boldsymbol{k}}+\delta\boldsymbol{\Omega}_{\lambda_{i}\boldsymbol{k}})\cdot e\boldsymbol{E}.
\end{aligned} \label{LPS}
\end{equation}
The band index is suppressed for simplicity unless stated otherwise. In the
zeroth order of $\bs{E}$ field, the layer polarization is given by $p_{i}^{0}(\bs{k})=-\partial
_{\lambda_{i}}\varepsilon=\langle u|\hat{p}_{i}|u\rangle$, where $\varepsilon$
is the band energy. 
This term describes the intrinsic interlayer charge distribution.
The linear order response, $p_{i}^{\text{L}}(\bs{k})=e\boldsymbol{E}\cdot
\boldsymbol{\Omega}_{\lambda_{i}\boldsymbol{k}}$, corresponds to an anomalous
layer polarization contributed by the Berry curvature $\boldsymbol{\Omega
}_{\lambda_{i}\boldsymbol{k}}=\partial_{\lambda_{i}}\boldsymbol{\mathcal{A}%
}-\partial_{\boldsymbol{k}}\mathfrak{\bm A}_{i}$ in $\left(
\boldsymbol{k},\lambda_{i}\right)  $ space, where $\boldsymbol{\mathcal{A}%
}=\langle u|i\partial_{\boldsymbol{k}}|u\rangle$ and $\mathfrak{\bm A}%
_{i}=\langle u|i\partial_{\lambda_{i}}|u\rangle$ are the $k$-space and
$\lambda_{i}$-space Berry connections, respectively. For the $n$-th band, this Berry curvature explicitly reads (set $\hbar=1$)%
\begin{equation}
\boldsymbol{\Omega}_{\lambda_{i}\boldsymbol{k}}^{n}=2{\operatorname{Im}}\sum_{m\neq n}\frac{p_{i}^{nm}\left(
\boldsymbol{k}\right)  \boldsymbol{v}^{mn}\left(  \boldsymbol{k}\right)
}{[\varepsilon_{n}\left(  \boldsymbol{k}\right)  -\varepsilon_{m}\left(
\boldsymbol{k}\right)  ]^{2}},~\label{eq:Omega_mix}
\end{equation}
whose numerator involves interband matrix elements of the multipole
and velocity operators. 
The second order ($E^2$) nonlinear response $p_{i}^{\text{NL}}(\bs{k})=\delta\bs{\Omega}_{\lambda_{i}\bs{k}}\cdot e\bs{E}-\partial_{\lambda_{i}}\delta\varepsilon$ contains two contributions. The first part originates from the correction of $\left(  \boldsymbol{k},\lambda_{i}\right)  $-space Berry curvature by $\bs{E}$ field, $\delta\boldsymbol{\Omega}_{\lambda
_{i}\boldsymbol{k}}=\partial_{\lambda_{i}}\boldsymbol{\mathcal{A}}%
^{E}-\partial_{\boldsymbol{k}}\mathfrak{\bm A}_{i}^{E}$,
where $\mathcal{A}_{a}^{E}=G_{k_a k_b}E_{b}$
and $\mathfrak{\bm A}_{i}^{E}=\boldsymbol{\mathcal{G}}_{\lambda_{i}\bs{k}}\cdot\boldsymbol{E}$
are the field induced $k$-space and $\lambda_{i}$-space Berry connections,
respectively, with \cite{Gao2014,Chen2023}%
\begin{align}
\boldsymbol{\mathcal{G}}_{\lambda_{i}\bs{k}}^{n}
& =  2e\mathrm{{\operatorname{Re}}}\sum_{m\neq n}\frac{p_{i}^{nm}\left(
	\boldsymbol{k}\right)  \boldsymbol{v}^{mn}\left(  \boldsymbol{k}\right)
}{[\varepsilon_{n}\left(  \boldsymbol{k}\right)  -\varepsilon_{m}\left(
	\boldsymbol{k}\right)  ]^{3}} \label{BCP_in_kl}\\
 G_{k_a k_b}^{n} & =  -2e\mathrm{{\operatorname{Re}}}
\sum_{m\neq n}\frac{v_{a}^{nm}\left(  \boldsymbol{k}\right)  v_{b}^{mn}\left(
	\boldsymbol{k}\right)  }{[\varepsilon_{n}\left(  \boldsymbol{k}\right)
	-\varepsilon_{m}\left(  \boldsymbol{k}\right)  ]^{3}}  \label{BCP_in_kk}
 \end{align}
being the corresponding BCPs. Here, summation over repeated Cartesian indices
($a,b$) is implied. 
The second part of $p_{i}^{\text{NL}}(\bs{k})$ is rooted in the second order
correction of the electron energy $\delta\varepsilon=eG_{k_a k_b}E_{a}E_{b}/2$~\cite{Xiao2022MBT}.
It is apparent that the anomalous layer polarization induced by in-plane $\bs{E}$ field is prominent around nearly degenerate regions of the bands, as a result of their band geometric nature.

Finally, the layer polarization density is obtained by
$p_{i}=\int[d\boldsymbol{k}]f\left(  \boldsymbol{k}\right)  p_{i}\left(
\boldsymbol{k}\right)$,
where $[d\boldsymbol{k}]=\sum_{n}d\boldsymbol{k}/(2\pi)^{2}$ and $f$ is the
distribution function. We focus on the intrinsic response determined
solely by the band structures, thus take the equilibrium Fermi distribution
$f_{0}$. The linear and nonlinear interlayer multipole generation can be obtained
from $p_{i}^{\text{L}}(\bs{k})$ and $p_{i}^{\text{NL}}(\bs{k})$ respectively. In the following, we assume a uniform ac driving field
$\boldsymbol{E}=\boldsymbol{E}^{0}\cos\omega t$ with frequency $\omega$, which facilitates observing the effect by harmonic generations of interlayer multipoles. It is noted that the proposed effect is of nonequilibrium
nature, the same as the intrinsic electrical spin generation
\cite{Kurebayashi2014,Zelezny2014} and intrinsic anomalous/valley Hall effect.
The nonequilibrium effects are evaluated in the uniform limit (or
transport limit) as usual electrical transport effects
\cite{Luttinger1964,Zhong2016}, in which the local electron density does not
change upon the application of the $\bs{E}$ field. 

\emph{Linear and nonlinear polarization responses.-} The linear interlayer
polarization response is described by
$p_{i}^{\text{L}}=\alpha_{ia}E_{a}$, where the linear response coefficient
\begin{equation}
\alpha_{ia}=e\int[d\boldsymbol{k}]f_{0}\Omega_{\lambda_{i}k_{a}}%
\end{equation}
is intrinsic to the band structure, free of scattering effects, and involves
only the unperturbed mixed-space Berry curvature over the occupied states. The form of $\alpha_{ia}$ characterizing the linear response is constrained by symmetry, detailed symmetry analysis are provided in Supporting Information. 

To the intrinsic second order ($E^2$), multipole generation takes the form of
$p_{i}^{\text{NL}}=\alpha_{iab}E_{a}^{0}E_{b}^{0}+\alpha_{iab}E_{a}^{0}E_{b}^{0}\cos2\omega t$,
which consists of both static and second harmonic components.
Here we focus on the second harmonic component that can be readily separated
from the possible intrinsic static polarization. Only the symmetric part of
the response tensor $\alpha_{i(ab)}=(\alpha_{iab}+\alpha_{iba})/2$
contributes, which reads
\begin{equation}
\alpha_{i(ab)}=\frac{e}{2}\int[d\boldsymbol{k}]f_{0}(\partial_{\lambda_{i}%
}G_{k_a k_b}-\partial_{k_{a}}\mathcal{G}_{\lambda_i k_b}-\partial_{k_{b}}\mathcal{G}_{\lambda_i k_a}).
\label{NL}%
\end{equation}
It contains the dipole moments of BCPs in the $\left(  \boldsymbol{k},\lambda
_{i}\right)  $ space, hence is a property of interlayer hybridized
electronic states.
Inside insulating gaps, it becomes $\alpha_{i(ab)}=e\int[d\boldsymbol{k}]f_{0}\partial_{\lambda_{i}}G_{k_a k_b}/2$, which only depends on the $k$-space BCP of
occupied states. While in metallic regimes, there are additional Fermi surface terms
quantified by the symmetric part of the dipole of mixed-space BCP. A recent study has highlighted the role of the antisymmetric part
of this generalized BCP dipole in the crossed nonlinear dynamical Hall effect
\cite{Chen2023}, a novel intrinsic nonlinear Hall effect unique to chiral nonmagnetic layered systems, whereas the result here unveils the connection of its symmetric part to nonlinear polarization responses.
In the metallic regime, both the Fermi sea and Fermi surface terms contribute significantly to the net response, while the values of individual contributions depend on the details of the system.
The form of $\alpha_{i(ab)}$ in a specific crystal class is also constrained by symmetry, as detailed in Supporting Information. 

\emph{Quantum metric contribution to the nonlinear responses.-} Expressions of the BCPs in Eqs.~(\ref{BCP_in_kl}) and (\ref{BCP_in_kk}) resemble that of the Fubini-Study metric (FSM)~\cite{provost1980riemannian}, which measures the `quantum distance' of parametrized eigenstates in projected Hilbert space, suggesting that the nonlinear multipole responses are closely connected to quantum metric. Indeed, the intrinsic nonlinear Hall effect has been shown to be related to the $k$-space quantum metric~\cite{Ahn2020,watanabe2021chiral,gao2023quantum,wang2023quantum}, whereas the crossed dynamical nonlinear Hall effect unique to chiral layered systems is connected to the mixed quantum metric in $(\bs{k},\lambda_{1})$ space \cite{Chen2023}. Here, we show that the quantum metric in $(\bs{k},\lambda_{i})$ space underlies the intrinsic nonlinear interlayer multipole response to in-plane field, as detailed in Supporting Information.
One can then split the BCPs into contributions from the FSMs and additional remote bands (ARB):
\begin{align}
G_{k_a k_b}^{n}&=-2e\frac{g_{ab}^{n}}{\varepsilon_{n \bar{n}}}+G_{k_a k_b}^{n,\text{ARB}},\\
\mc{G}_{\lambda_i k_a}^{n}&=-2e\frac{\mathfrak{g}_{ia}^{n}}{\varepsilon_{n \bar{n}}}+\mc{G}_{\lambda_i k_a}^{n,\text{ARB}},
\end{align}
where $g^{n}_{ab}$ and $\mathfrak{g}^{n}_{ia}$ are FSMs in the $k$ and $(\bs{k},\lambda_i)$ spaces, respectively. $\varepsilon_{n \bar{n}}=\varepsilon_{n}-\varepsilon_{\bar{n}}$ is the energy difference between $\varepsilon_{n}$ and its nearest neighbor $\varepsilon_{\bar{n}}$.
The nonlinear response coefficients can thus be decomposed into the FSM and ARB contributions, $\alpha_{i(ab)} = \alpha^{\text{FSM}}_{i(ab)} + \alpha^{\text{ARB}}_{i(ab)}$, where
\begin{eqnarray}
    \alpha^{\text{FSM}}_{i(ab)} = -e^2 \int [d\boldsymbol{k}] f_0 (\partial_{\lambda_i}\frac{g^n_{ab}}{\varepsilon_{n \bar{n}}} - \partial_{k_a}\frac{\mathfrak{g}^n_{i b}}{\varepsilon_{n \bar{n}}} - 
    \partial_{k_b}\frac{\mathfrak{g}^n_{i a}}{\varepsilon_{n \bar{n}}}) \label{Eq:FSM_to_nonlinear}
\end{eqnarray}
is related to the FSM contributions to BCP.
In the cases where there are only two bands, all the terms associated with ARB vanish, thus FSM is the only contribution.
In multi-band situations, the FSM contribution dominates if band $n$ is very close to another band but well-separated from the rest. Notice that the last two terms in Eq.~(\ref{Eq:FSM_to_nonlinear}) require Fermi surface to survive, thus one can focus on the contribution of the $k$-space FSM, $g^n_{ab}$, in the insulating regime.

\begin{figure}[ptb]
\includegraphics[width=0.7\textwidth]{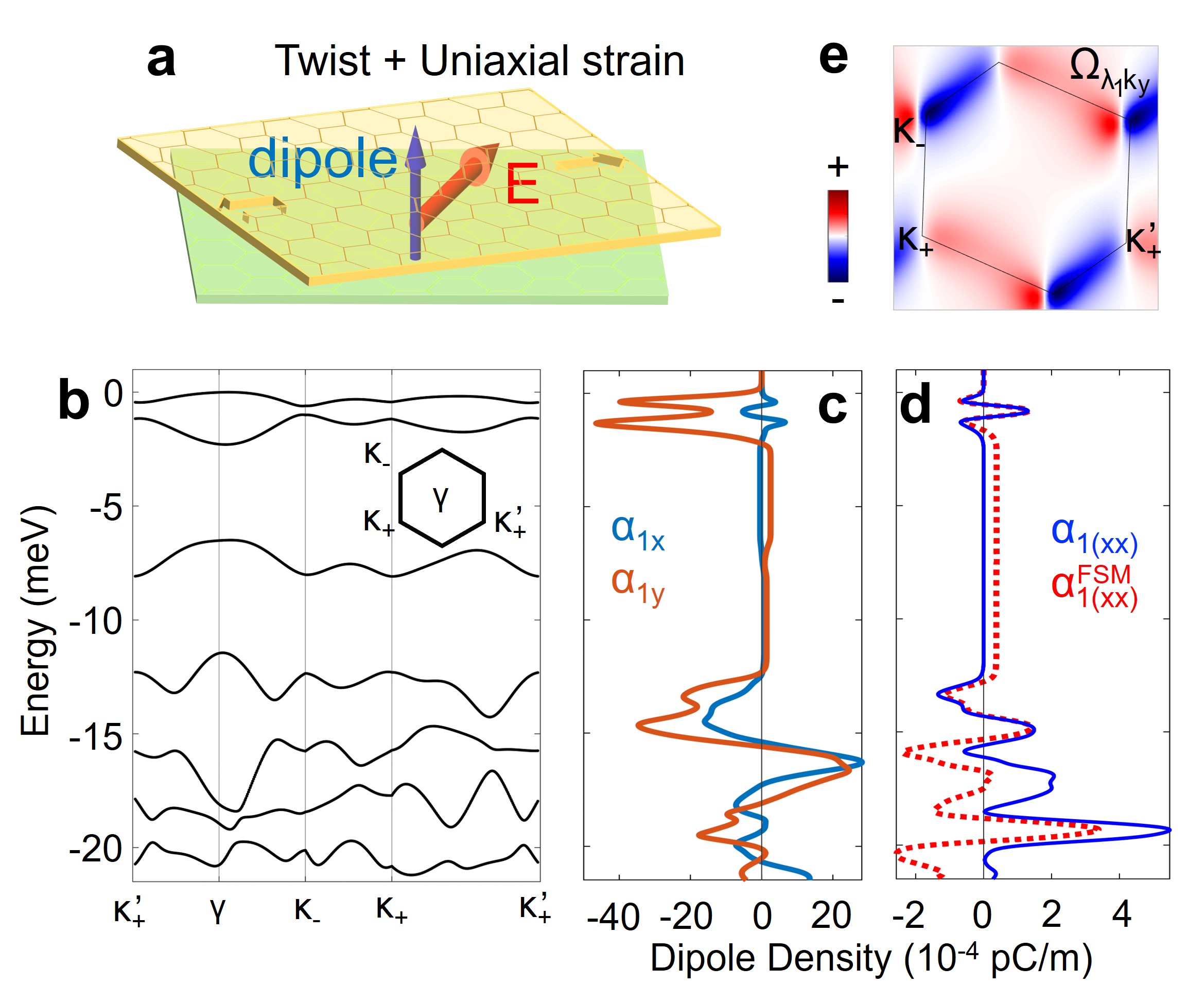} \caption{(a) Schematics of
the anomalous dipole polarization in bilayer MoTe$_{2}$ with twisting and uniaxial
heterostrain. Low-energy valence bands (b), and dipole density as a function of Fermi energy from linear (c) and nonlinear (d) effects by electric field of $10^{4}$V/m. Twisting angle $\theta=1.2^{\circ}$ and
heterostrain $\epsilon=0.3\%$. In (d), the blue and red dashed curves represent the exact result and Fubini-Study metric contribution, respectively. (e) \textit{k}-space distribution of $\Omega
_{\lambda_{1} k_{y}}$ in the first miniband. }%
\label{fig_mote2}%
\end{figure}

\emph{Interlayer polarization in twisted bilayers.-} To illustrate
the features of the linear polarization effect and the underlying Berry
curvature $\bs{\Omega}_{\lambda_{i}\bs{k}}$, we take the superlattice of twisted bilayer MoTe$_{2}$ with uniaxial strain~\cite{qiao2018twisted,huder2018electronic,androulidakis2020tunable,gao2021heterostrain,Tara2022} as an
example (Fig.~\ref{fig_mote2}a). The relative twist of the two layers breaks all the mirror symmetries. Yet the system can have rotation symmetries, which forbid linear dipole generation. An uniaxial strain, which we take to be along the zigzag direction of the top layer (Fig.~\ref{fig_mote2}a), 
can break all the remaining rotational symmetries, allowing the linear dipole generation to emerge.

To study the layer polarization effect quantitatively, we use the continuum model~\cite{wu2019topological}, and incorporate the effects of strain~\cite{Bi2019,ZhaiPRM2020,fang2018electronic}.
Figure~\ref{fig_mote2}b shows the low-energy valence band structures of
twisted MoTe$_{2}$ at twist angle $\theta=1.2^{\circ}$ and $0.3\%$ strain. 
Results for larger twist angles are presented in Fig.~S1 of the Supporting Information.
The uniaxial strain distorts the mini-BZ (mBZ) and shifts the Dirac point of the top/bottom layer away from the $\bs{\kappa_{-}} / \bs{\kappa_{+}}$ corner.
The Berry curvature $\boldsymbol{\Omega}_{\lambda_{1}\boldsymbol{k}}$ is concentrated at band near-degeneracy regions, as is shown in Fig.~\ref{fig_mote2}e for its \textit{k}-space distribution on the first valence band. Figure~\ref{fig_mote2}c shows the dipole density produced by the linear response as a function of chemical potential.
As the effect favors nearly degenerate regions, each energy band that is in close proximity to adjacent bands is associated with a prominent peak. The third band contributes negligibly as it is well separated from its neighbors.
The presence of interlayer dipole can be manifested as a voltage drop across the layers, which can be estimated as $\delta V=p^{\text{L}}/\varepsilon_{0}$ with $\varepsilon_{0}$ being the vacuum permittivity~\cite{Gao2020}. Under a moderate electric field of $10^{4}$ V/m, the induced linear dipole density can reach $p^{\text{L}}\sim O(10^{-3})$~pC/m, which corresponds to an interlayer voltage of $\delta V\sim O(0.1)$ mV. Such polarization response can reach over a tenth of the intrinsic polarization of the system (see Supporting Information).  Later we will show that larger interlayer voltage reaching a few mV can be achieved in trilayers with appropriate translations.

In addition to linear response, twisted bilayer MoTe$_{2}$ with uniaxial strain also supports the nonlinear dipole generation. The blue curve in Fig.~\ref{fig_mote2}d shows $\alpha_{1(xx)}$ as an example. Especially, at low energies there is a sharp
peak centered in the gap between the first two bands, which corresponds to the Fermi sea contribution in Eq.~(\ref{NL}), i.e., $\alpha_{i(ab)}=\frac{e}{2}\int[d\boldsymbol{k}]f_{0}\partial_{\lambda_{i}}G_{k_a k_b}$. Due to the
flat bands with larger density of states and small local gaps in the system, the induced dipole density (thus the interlayer voltage) by the nonlinear effect is comparable to that from the linear response under a moderate field of $10^{4}$ V/m (cf. Figs.~\ref{fig_mote2}c and d).

As discussed earlier, the FSMs, $g_{ab}$ and $\mathfrak{g}_{ia}$, contribute to the nonlinear interlayer response [Eq.~(\ref{Eq:FSM_to_nonlinear})]. The red dashed curve in Fig.~\ref{fig_mote2}d shows the FSM contribution. Within the energy window of the first two bands, the nonlinear response is almost solely contributed by the FSM. 
This is well expected as these two bands are close to each other, while they are well separated from other bands, thus the remote band contribution is negligible. A similar situation can also be identified in the energy range of the fourth band. Therefore, such energy windows are ideal for exploring the effects of the FSM. Especially, one can tune the chemical potential into the energy gaps to focus only on the contribution of the $k$-space FSM $g_{ab}$.
As the chemical potential is tuned further, e.g., below $-15$ meV, the bands become more crowded. In such cases, in addition to the closest band $\bar{n}$, the remote bands will also contribute significantly to the layer polarization on band $n$, hence the deviation between the exact result and FSM contribution increases.

\begin{figure}
	\includegraphics[width=0.6\textwidth]{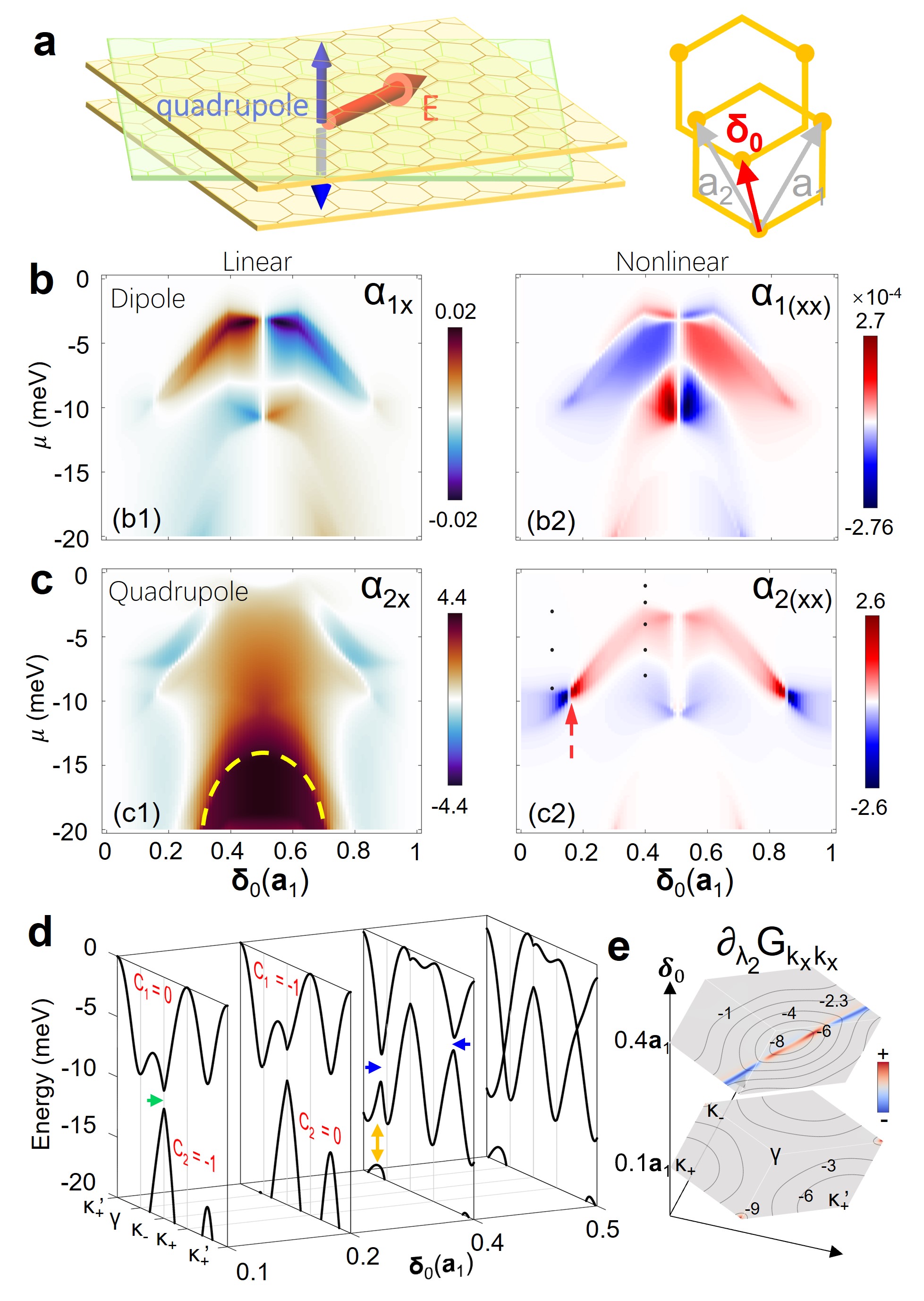} 
	\caption{$\mathbf{a}$ Schematics of the anomalous quadrupole response in twisted trilayer MoTe$_{2}$. $\bs{\delta}_{0}$ parameterises the top layer displacement with respect to the bottom layer. $\mathbf{b}$ Left and right panels show dipole densities (units: pC/m) from the linear and nonlinear effects, respectively. Only $x$ and $xx$ components are shown, and the other components exhibit similar feature and magnitude.
	$\mathbf{c}$ Similar plots for quadrupole density (units: $10^{-13}$ pC). The red arrow marks the topological phase transition point. $\mathbf{d}$ Dispersion in the $K$ valley along the high-symmetry line and band Chern numbers. $\mathbf{e}$ Momentum space distribution of the band geometric quantity $\partial_{\lambda_{2}}G_{k_x k_x}$ in the first valence band for the nonlinear quadrupole response. Two examples at $\boldsymbol{\delta}_0=0.1\mathbf{a}_1$ and $\boldsymbol{\delta}_0=0.4\mathbf{a}_1$ are shown. Twist angle $\theta=3^\circ$, an electric field of $10^4$ V/m is considered, and 0.3 meV thermal broadening is used in the calculations.}
	\label{fig_displacement}%
\end{figure}

\emph{Interlayer polarization in twisted trilayers.-} Much more abundant interlayer polarization responses will emerge when the layer number is increased to three. 
Here we consider twisted MoTe$_2$ trilayer as an example, and assume that the top and bottom layers are parallel to maintain the moir\'e periodicity. Two parameters can be used to specify the structural geometry (Fig.~\ref{fig_displacement}a): (i) the twist angle $\theta$ of the outer layers relative to the middle layer, (ii) the translation $\bs{\delta}_0$ of the top layer with respect to the bottom layer. When outer layers are fully aligned ($\boldsymbol{\delta}_0 = 0$), the trilayer possesses $C_{3h}$ point group, which only allows nonlinear quadrupole generation, this special case is demonstrated in Supporting Information.

To illustrate the effects of translation, we consider $\bs{\delta}_0 = c \bs{a}_1 $ with $c\in [0,1]$ in a $3^\circ$ twisted trilayer as an example (Fig.~\ref{fig_displacement}a), which is chosen along a zigzag direction of the monolayer.
A nonzero $\bs{\delta}_0$ in general breaks $\mathcal{M}_{z}$ and $C_3$ symmetries, thus dipole and quadrupole generation can emerge simultaneously with both linear and nonlinear responses.
Figures~\ref{fig_displacement}b and c show various layer polarization responses in variation of translation and chemical potential ($\mu$). 
First we summarize a few interesting trends identified from the figures.
(i) The dipole (quadrupole) responses show anti-symmetric (symmetric) profiles with respect to the vertical line at $\bs{\delta_0} = 0.5\bs{a}_1$. Therefore, it suffices to focus on half of the plots.
(ii) The pronounced features in Figs.~\ref{fig_displacement} (b1), (b2) and (c2) fall in similar regions of the parameter space $(\mu,\,\bs{\delta}_{0})$. These regions, as shown later, correspond to the energy windows enclosing band near-degeneracy at different $\bs{\delta}_0$.
(iii) The responses show opposite signs on the two sides of $\bs{\delta_0} \approx 0.15\bs{a}_1$ around $\mu\approx-10$ meV [see, e.g., red arrow in Fig.~\ref{fig_displacement}(c2)]. As will be shown shortly, this corresponds to a topological phase transition point, implying the multipole generation might be used for probing topological properties of the system.

Now we offer more insights on the above observations. Point (i) is rooted in a `mirror symmetry' of the system: a backward translation $-\bs{\delta}_0$ of the top layer is equivalent to a forward translation $\bs{\delta}_0$ of the bottom layer, whose geometry is $\mc{M}_z$-symmetric to that by shifting $\bs{\delta}_0$ of the top layer. The $\mc{M}_z$ operation inverts the dipole and leaves the quadrupole unaffected, thus rendering Figs.~\ref{fig_displacement}b and c anti-symmetric and symmetric, respectively. 
Points (ii) and (iii) are closely connected to the evolution of the band structures with $\bs{\delta}_{0}$.
Figure~\ref{fig_displacement}d presents low-energy valence bands within $[-20,\,0]$ meV for a few values of $\bs{\delta}_{0}$. 
When $\bs{\delta}_{0}=0.1\bs{a}_1$, the first two bands are nearly degenerate around $\bs{\kappa_{-}}$ corners of the mBZ (see the green arrow in Fig.~\ref{fig_displacement}d) and a small global gap exists. 
The gap closes and reopens as $\bs{\delta}_{0}$ is evolved from $0.1\bs{a}_1$ to $0.2\bs{a}_1$. In this process, a topological phase transition occurs with the Chern numbers of the first two bands interchanged (compare first two plots of Fig.~\ref{fig_displacement}d), which is accompanied by a quadrupole sign flip in both linear and nonlinear responses [see e.g., red arrow in Fig.~\ref{fig_displacement}(c2)]. The remarkable resemblance between Eq.~(\ref{eq:Omega_mix}) that determines the linear response and the $k$-space Berry curvature implies that there might exist an intricate connection between the topological transition and the sign reversal of multipole generation. However, systematic future studies are needed to explore this. The gap reduction around the transition region greatly enhance the responses. 
As $\bs{\delta}_{0}$ is increased further, the small-gap regions move away from $\bs{\kappa_{-}}$ (see the blue arrows in Fig.~\ref{fig_displacement}d at $\bs{\delta_0} = 0.4\bs{a}_1$) and become elongated 1D-like on the $k_x$-$k_y$ plane. 
Finally, at $\bs{\delta_0} = 0.5\bs{a}_1$, the first two bands touch again (See Supporting Information). 

Since the band geometric quantities favor small energy separations, they can be used to visualize the evolution of band near-degeneracy. Figure~\ref{fig_displacement}e shows the distribution of $\partial_{\lambda_{2}}G_{k_x k_x}$ on the first valence band for $\bs{\delta_0} = 0.1\bs{a}_1$ and $0.4\bs{a}_1$. As expected, it is localized around the three $\bs{\kappa_{-}}$ corners of the mBZ when $\bs{\delta_0} = 0.1\bs{a}_1$, while it becomes 1D-like at $\bs{\delta_0} = 0.4\bs{a}_1$. This geometric quantity contributes to the nonlinear quadrupole generation in Fig.~\ref{fig_displacement}(c2). We have selected a few energies, which are plotted as equal-energy contours in Fig.~\ref{fig_displacement}e and as black dots in Fig.~\ref{fig_displacement}(c2). By comparing the two figures, one confirms that the regions with strong multipole responses indeed correlate with the band near-degeneracy.

Lastly we comment on the yellow dashed dome in Fig.~\ref{fig_displacement}(c1). This feature corresponds to the energy gap separating the second and third bands (see e.g., yellow arrow in Fig.~\ref{fig_displacement}d at $\bs{\delta_0} = 0.4\bs{a}_1$). This insulating region is advantageous for exploring the effects due to $\Omega_{\lambda_2 \bs{k}}$ and $\partial_{\lambda_2} G_{k_a k_b}$.

\begin{acknowledgement}
This work is supported by the Research Grant Council of Hong Kong (AoE/P-701/20, HKU SRFS2122-7S05, 17306819), the New Cornerstone Science Foundation, and the Start-up Research Grant at University of Macau (SRG2023-00033-IAPME).
\end{acknowledgement}


\begin{suppinfo}
The Supporting Information is available free of charge at http://pubs.acs.org.

S1. Interpretation of the coupling term in the main text. S2. Derivation of Eq. (1) in the main text. S3. Explicit expression of $\partial_{\lambda_i}G_{k_a k_b}$. S4. Symmetry constraint of linear and nonlinear effects. S5. Quantum metric contribution of intrinsic nonlinear layer polarization. S6. Continuum model of twisted bilayer MoTe$_2$. S7. Dipole responses at larger twist angles of a twisted bilayer. S8. Continuum model of twisted trilayer MoTe$_2$ and layer polarization in mirror-symmetric case. S9. Comparison of polarization response with the intrinsic layer polarization. 
\end{suppinfo}

\bibliography{LEE_ref}

\end{document}


\tableofcontents

\section{Interpretation of the coupling term in the main text}
The coupling term introduced in the main text can be interpreted as the electrostatic energy. Intuitively, one can write the electrostatic energy as $W = -\sum_l \rho_l \Phi_l$, which consists of the charge density $\rho_l$ on layer $l$ and the corresponding electric potential $\Phi_l$. It can be rewritten in a multipole expansion manner: 
\begin{eqnarray}
W = -M\cdot\bar{\Phi}-D\cdot E_{\perp}-Q\cdot E'_{\perp}-\cdots,
\label{formula_totalenergy}
\end{eqnarray}
where in each term the first letter denotes the interlayer multipole moment (e.g., monopole, dipole, quadrupole, etc), and the second letter represents the associated coupling `field' (e.g., average potential, electric field, gradient of electric field, etc).
As examples, let us focus on bilayers and trilayers (Fig. 1 of the main text).
In bilayers, one can readily identify the monopole (charge density) $M=\rho_1+\rho_2$ and average potential $\bar{\Phi} = (\Phi_1 + \Phi_2)/2$. The dipole moment reads $D=d(\rho_2-\rho_1)/2$ with average out-of-plane electric field $E_{\perp} = (\Phi_2 - \Phi_1)/d$, where $d$ is the interlayer distance. One can verify with these expressions that $W_{\text{bilayer}}=-M\cdot\bar{\Phi}-D\cdot E_{\perp}=-\rho_1\Phi_1-\rho_2\Phi_2$. Similarly, in trilayers one finds $M=\rho_1+\rho_2+\rho_3$ and $\bar{\Phi} = (\Phi_1 + \Phi_2 + \Phi_3)/3$ for the monopole part, and $D=d(\rho_3-\rho_1)$ with $E_{\perp} = (\Phi_3 - \Phi_1)/(2d)$ for the dipole contribution. Additionally, there exists a quadrupole moment, which is found to be $Q=d^2(\rho_3-2\rho_2+\rho_1)/6$ by approximating the gradient of the $E_{\perp}$ field as $E'_{\perp} = (\Phi_3 -2\Phi_2 +\Phi_1)/d^2$.

In a crystal, these multipoles are contributed by electrons characterized by Bloch states, whose periodic part is denoted as $\ket{u_{n\bs{k}}}$ with $n$ and $\boldsymbol{k}$ being the band index and crystal momentum, respectively. Each state's contribution to the multipole can be evaluated as $p_i(n,\bs{k}) = \braket{u_{n\bs{k}}|\hat{p}_{i}|u_{n\bs{k}}}$, where we have used $\hat{p}_{i=0,1,2,\cdots}$ to represent the multipole operators in ascending order. By using the operators in Fig.~1 of the main text, one finds expressions of the multipoles consistent with the results in the above.

In addition, there exists intrinsic relationship between the layer multipole moments and the layer
charge distribution.
\begin{eqnarray}
    \begin{pmatrix}
        \rho_1 \\ \rho_0
    \end{pmatrix}
     &=& p_0
 \begin{pmatrix}
     1 \\ 1
 \end{pmatrix}
  + \frac{p_1}{d}
  \begin{pmatrix}
      1 \\ -1
  \end{pmatrix}, \\
      \begin{pmatrix}
        \rho_2 \\ \rho_1 \\ \rho_0
    \end{pmatrix}
     &=& p_0
 \begin{pmatrix}
     1 \\ 1 \\ 1
 \end{pmatrix}
  + \frac{2p_1}{d}
  \begin{pmatrix}
      1 \\ 0 \\ -1
  \end{pmatrix}
    + \frac{p_2}{d^2}
  \begin{pmatrix}
      1 \\ -2 \\ 1
  \end{pmatrix}.
\end{eqnarray}
The above relation shows that the layer multipole moment dictates the components of the corresponding layer charge
distribution (Fig.~1 in the maintext).

\section{Derivation of Eq. (1) in the main text}

The $p_{i}^{\text{L}}(\bs{k})$ and $p_{i}^{\text{NL}}(\bs{k})$ of the Eq.~(1) in the main text are the \textit{anomalous} multipole carried by a particular Bloch electron. The terminology `anomalous' follows the similar origin as the well known anomalous velocity of Bloch electrons~\cite{Xiao2010}.
The origin of this term can be understood intuitively as follows. The applied in-plane electric field $\bm E$ polarizes the periodic part of the Bloch state in the form of 
\begin{equation}
    |u_{n}\rangle \rightarrow |u_{n}\rangle + \sum_{n' \neq n} \frac{e\bm E \cdot \bm{\mathcal{A}}_{n'n}}{\varepsilon_n - \varepsilon_{n'}}|u_{n'}\rangle,
\end{equation}
where $\bm{\mathcal{A}}_{n'n}$ is the $k$-space interband Berry connection. The expectation value of the charge multipole operator on this field-modified state, in the first order of $E$ field, gives rise to the anomalous multipole. This result can be confirmed by systematic derivation of semiclassical wave packet theory \cite{Dong2020}. Such a wave packet theory has also been extended to the second order of electric field, where the $E$ field corrects both the anomalous multipole and the electron energy in the first term in the second line of Eq. (1) in the main text (see the systematic derivation in the Supplemental Material~\cite{Xiao2022MBT}).

\section{Explicit expression of $\partial_{\lambda_i}G_{k_ak_b}$}
The $\lambda_i$-derivative of intralayer BCP, $\partial_{\lambda_i}G_{k_ak_b}$, reads:
\begin{align}
\partial_{\lambda_i}G_{k_ak_b}^{n}    =&-2e\hbar^{2}\mathrm{{\operatorname{Re}}}%
\sum_{m\neq n}\frac{3\left(  p^{n}_i-p^{m}_i\right)  v_{a}^{nm}v_{b}^{mn}%
}{(\varepsilon_{n}-\varepsilon_{m})^{4}}
  +2e\hbar^{2}\mathrm{{\operatorname{Re}}}\sum_{m\neq n}\sum_{\ell\neq
n}\frac{p^{n\ell}_i\left(  v_{a}^{\ell m}v_{b}^{mn}+v_{b}^{\ell m}v_{a}%
^{mn}\right)  }{\left(  \varepsilon_{n}-\varepsilon_{l}\right)  (\varepsilon
_{n}-\varepsilon_{m})^{3}}\nonumber\\
&  +2e\hbar^{2}\mathrm{{\operatorname{Re}}}\sum_{m\neq n}\sum_{\ell\neq
m}\frac{p^{m\ell}_i\left(  v_{a}^{\ell n}v_{b}^{nm}+v_{b}^{\ell n}v_{a}%
^{nm}\right)  }{\left(  \varepsilon_{m}-\varepsilon_{l}\right)  (\varepsilon
_{n}-\varepsilon_{m})^{3}}.
\end{align}
One observes that this band quantity can be nonzero only if the layer pseudospin is not
conserved, which requires interlayer hybridized electronic wave functions.

\section{Symmetry Constraint of linear and nonlinear effects}
The form of $\alpha_{ia}$ characterizing the linear response is constrained by symmetry. First, the dipole ($i=1$) and quadrupole ($i=2$) responses both are prohibited by any rotation symmetry about the $z$ direction. Furthermore, the dipole generation is allowed by
the inversion symmetry, but prohibited by the horizontal mirror plane $\mathcal{M}_{z}$. Whereas the quadrupole generation is prohibited by the
inversion symmetry, but allowed by the horizontal mirror plane $\mathcal{M}_{z}$. The detailed results of symmetry analysis are shown in Table \ref{LR}.
One sees that the largest point groups supporting the linear dipole and quadrupole effects are $C_{2h}$ and $C_{2v}$, respectively.

\begin{table}[h]
\caption{{}Crystal classes pertaining to 2D materials in which the linear
dipole and quadrupole generations are allowed and the corresponding response
structures.}%
\label{LR}%
\begin{tabular}
[b]{cc}\hline\hline
\ point groups & \ layer dipole\\\hline
$C_{1},S_{2}$ & $\alpha_{1x}E_{x}+\alpha_{1y}E_{y}$\\
$C_{1v}$ ($\mathcal{M}_{x}$),$C_{2}$ ($C_{2}^{x}$ axis)$,C_{2h}=C_{1v}\otimes
C_{2}$ & $\alpha_{1y}E_{y}$\\\hline\hline
\ point groups & \ layer quadrupole\\\hline
$C_{1},C_{1h}$ & $\alpha_{2x}E_{x}+\alpha_{2y}E_{y}$\\
$C_{1v}$ ($\mathcal{M}_{x}$) & $\alpha_{2y}E_{y}$\\
$C_{1v}$ ($\mathcal{M}_{y}$),$C_{2}$ ($C_{2}^{x}$ axis)$,C_{2v}=C_{1v}\otimes
C_{2}$ & $\alpha_{2x}E_{x}$\\\hline\hline
\end{tabular}
\end{table}

\begin{table}[h]
\caption{{}Crystal classes pertaining to 2D materials in which the second
order nonlinear dipole generation is allowed and the corresponding
response structures. }
\begin{tabular}
[b]{cc}\hline\hline
\ \ point groups \ \  & \ layer dipole\ \\\hline
& \\
$C_{1},C_{2}$\ $(C_{2}^{z}$ axis$)$ & $\alpha_{1xx}E_{x}^{2}+2\alpha
_{1(xy)}E_{x}E_{y}+\alpha_{1yy}E_{y}^{2}$\\
& \\
$C_{1v},C_{2v}$\ $(C_{2}^{z}$ axis$)$ & $\alpha_{1xx}E_{x}^{2}+\alpha
_{1yy}E_{y}^{2}$\\
& \\
$C_{3},C_{4},C_{6},C_{3v},C_{4v},C_{6v},$ & $\alpha_{1xx}E^{2}$\\
& \\
$C_{2}$ $(C_{2}^{x}$ axis$),D_{2}$\  & $2\alpha_{1(xy)}E_{x}E_{y}$\\
& \\
$D_{2d}$ ($\mathcal{M}_{x}$ and $\mathcal{M}_{y}$) & $\alpha_{1xx}(E_{x}%
^{2}-E_{y}^{2})$\\
& \\
$S_{4}\ $ & $\alpha_{1xx}(E_{x}^{2}-E_{y}^{2})+2\alpha_{1(xy)}E_{x}E_{y}%
$\\\hline\hline
\end{tabular}
\label{NLR}%
\end{table}

The form of $\alpha_{i(ab)}$ characterizing the nonlinear response in a specific crystal class is also constrained
by symmetry. The nonlinear interlayer dipole generation is prohibited by the inversion, $\mathcal{M}_{z}$, and the combination of an in-plane $C_{2}$ axis and
a more than twofold rotation axis in the $z$ direction. The resultant crystal
classes and the supported response structures are shown in Table \ref{NLR}.
In contrast, the nonlinear interlayer quadrupole generation is much less symmetry
constrained, and is in fact allowed in all crystal classes pertaining to 2D
materials. In particular, in point groups containing a more than twofold
rotation or roto-reflection, such as $D_{3}$, $C_{3h}$, $D_{3h}$, $D_{3d}$ and
$S_{4}$, and in point groups containing three twofold rotation axes
perpendicular to each other, including $D_{2}$, $D_{2h}$, and $D_{2d}$, the
nonlinear quadrupole response takes the isotropic form of $\alpha_{2xx}(E^{0})^{2}$, independent of the direction of the electric field.

\section{Quantum metric contribution of intrinsic nonlinear layer polarization}

One can relate the Berry connection polarizaibility (BCP) in $k$-space and $(\boldsymbol{k},\lambda_i)$-space with the quantum metric in the corresponding spaces:
\begin{eqnarray}
    G^n_{k_a k_b} = -2e\sum_{n_1 \neq n} \frac{g^{nn_1}_{ab}}{\varepsilon_n - \varepsilon_{n_1}},
\end{eqnarray}
and
\begin{eqnarray}
    \mathcal{G}^n_{\lambda_i k_b} = -2e\sum_{n_1 \neq n} \frac{\mathfrak{g}^{nn_1}_{ia}}{\varepsilon_n - \varepsilon_{n_1}}.
\end{eqnarray}
The numerator of $G_{k_ak_b}^{n}$:
\begin{equation}
g_{ab}^{nn_{1}}=\mathrm{{\operatorname{Re}}}\left[  \langle u_{n}%
|i\partial_{k_{a}}|u_{n_{1}}\rangle\langle u_{n_{1}}|i\partial_{k_{b}}%
|u_{n}\rangle\right]  =\mathrm{{\operatorname{Re}}}\left[  \langle
\partial_{k_{a}}u_{n}|u_{n_{1}}\rangle\langle u_{n_{1}}|\partial_{k_{b}}%
u_{n}\rangle\right]
\end{equation}
is the $\boldsymbol{k}$-space quantum metric for a pair of bands $n$ and
$n_{1}$, whereas the numerator of $\mathcal{G}_{\lambda_i k_a}^{n}$:%
\begin{equation}
\mathfrak{g}_{ia}^{nn_{1}}=\mathrm{{\operatorname{Re}}}\left[  \langle
u_{n}|i\partial_{\lambda_{i}}|u_{n_{1}}\rangle\langle u_{n_{1}}|i\partial
_{k_{a}}|u_{n}\rangle\right]  =\mathrm{{\operatorname{Re}}}\left[
\langle\partial_{\lambda_{i}}u_{n}|u_{n_{1}}\rangle\langle u_{n_{1}}%
|\partial_{k_{a}}u_{n}\rangle\right]
\end{equation}
is the quantum metric in $\left(  \boldsymbol{k},\lambda_{i}\right)  $
space for a pair of bands. These two quantities are gauge invariant and related to
the Fubini-Study quantum metric in $\left(  \boldsymbol{k},\lambda_{i}\right)
$ space as%
\begin{align}
g_{ab}^{n}  &  =\sum_{n_{1}\neq n}g_{ab}^{nn_{1}}=\mathrm{{\operatorname{Re}}%
}\langle\partial_{k_{a}}u_{n}|\left(  1-|u_{n}\rangle\langle u_{n}|\right)
|\partial_{k_{b}}u_{n}\rangle,\\
\mathfrak{g}_{ia}^{n}  &  =\sum_{n_{1}\neq n}\mathfrak{g}_{ia}^{nn_{1}%
}=\mathrm{{\operatorname{Re}}}\langle\partial_{\lambda_{i}}u_{n}|\left(
1-|u_{n}\rangle\langle u_{n}|\right)  |\partial_{k_{a}}u_{n}\rangle.
\end{align}
Here $g_{ab}^{n}$ is the $\boldsymbol{k}$-space\ Fubini-Study metric and
$\mathfrak{g}_{ia}^{n}$ is the mixed-space Fubini-Study metric. They together
describe the infinitesimal distance of quantum states in the parameter space
spanned by $\boldsymbol{k}$ and $\lambda_{i}$.

$G_{k_ak_b}^{n}$ and $\mathcal{G}_{\lambda_i k_a}^{n}$ can be decomposed into contributions from the
Fubini-Study metric (FSM) and additional remote bands (ARB):%
\begin{eqnarray}
G_{k_a k_b}^{n} &=& -2e\frac{g_{ab}^{n}}{\varepsilon_n - \varepsilon_{\bar{n}}}+G_{k_a k_b}^{n,\text{ARB}},\\
\mathcal{G}_{\lambda_i k_a}^{n} &=& -2e\frac{\mathfrak{g}_{ia}^{n}}{\varepsilon_n - \varepsilon_{\bar{n}}}+\mathcal{G}_{\lambda_i k_a}^{n,\text{ARB}},
\end{eqnarray}
where
\begin{eqnarray}
G_{k_a k_b}^{n,\text{ARB}}  =  -2e\sum_{n_{1}\neq n,\bar{n}%
}g^{nn_{1}}_{ab}\frac{\varepsilon_{n_{1}}-\varepsilon_{\bar{n}}}{\left(
\varepsilon_{n}-\varepsilon_{n_{1}}\right)  \left(  \varepsilon_{n}%
-\varepsilon_{\bar{n}}\right)  },\nonumber\\
\mathcal{G}_{\lambda_i k_a}^{n,\text{ARB}}  =  -2e\sum_{n_{1}\neq n,\bar{n}%
}\mathfrak{g}^{nn_{1}}_{ia}\frac{\varepsilon_{n_{1}}-\varepsilon_{\bar{n}}}{\left(
\varepsilon_{n}-\varepsilon_{n_{1}}\right)  \left(  \varepsilon_{n}%
-\varepsilon_{\bar{n}}\right)  },
\end{eqnarray}
with $\bar{n}$ being the band whose energy is closest to $n$. Accordingly, the
response coefficient of intrinsic nonlinear layer polarization can be decomposed into
\begin{equation}
\alpha=\alpha^{\text{FSM}}+\alpha^{\text{ARB}}.
\end{equation}
Here the FSM term reads%
\begin{align}
\alpha^{\text{FSM}}_{i(ab)} &  =-e^{2}\sum_{n}\int\frac{d\boldsymbol{k}}{\left(
2\pi\right)  ^{2}}(f_{0}\partial_{\lambda_{i}}\frac{g_{ab}^{n}}{\varepsilon
_{n}-\varepsilon_{\bar{n}}}+f_{0}^{\prime}\hbar v_{a}\frac{\mathfrak{g}%
_{ib}^{n}}{\varepsilon_{n}-\varepsilon_{\bar{n}}}+f_{0}^{\prime}\hbar
v_{b}\frac{\mathfrak{g}_{ia}^{n}}{\varepsilon_{n}-\varepsilon_{\bar{n}}%
})\nonumber\\
&  =-e^{2}\sum_{n}\int\frac{d\boldsymbol{k}}{\left(  2\pi\right)  ^{2}}\left[
f_{0}\frac{\partial_{\lambda_{i}}g_{ab}^{n}}{\varepsilon_{n}-\varepsilon
_{\bar{n}}}+f_{0}\frac{p_{i}^{n}-p_{i}^{\bar{n}}}{\left(  \varepsilon
_{n}-\varepsilon_{\bar{n}}\right)  ^{2}}g_{ab}^{n}+f_{0}^{\prime}\frac{\hbar
v_{a}}{\varepsilon_{n}-\varepsilon_{\bar{n}}}\mathfrak{g}_{ib}^{n}%
+f_{0}^{\prime}\frac{\hbar v_{b}}{\varepsilon_{n}-\varepsilon_{\bar{n}}%
}\mathfrak{g}_{ia}^{n}\right]
\end{align}
where
\begin{align}
\partial_{\lambda_{i}}g_{ab}^{n}  =\,&\hbar^{2}\operatorname{Re}\sum_{n_{1}\neq
n}\frac{2\left(  p_{i}^{n}-p_{i}^{n_{1}}\right)  v_{a}^{nn_{1}}v_{b}^{n_{1}n}%
}{\left(  \varepsilon_{n}-\varepsilon_{n_{1}}\right)  ^{3}}
  -\hbar^{2}\operatorname{Re}\sum_{n_{1}\neq n}\sum_{n_{2}\neq n}
\frac{p_{i}^{nn_{2}}(v_{a}^{n_{2}n_{1}}v_{b}^{n_{1}n}+v_{b}^{n_{2}n_{1}}v_{a}^{n_{1}n})}{\left(  \varepsilon
_{n}-\varepsilon_{n_{1}}\right)  ^{2}\left(  \varepsilon_{n}-\varepsilon
_{n_{2}}\right)  }  \nonumber\\
&  -\hbar^{2}\operatorname{Re}\sum_{n_{1}\neq n}\sum_{n_{2}\neq n_{1}}
\frac{p_{i}^{n_{1}n_{2}}(v_{a}^{n_{2}n}v_{b}^{nn_{1}} + v_{b}^{n_{2}n}v_{a}^{nn_{1}})}{\left(  \varepsilon
_{n}-\varepsilon_{n_{1}}\right)  ^{2}\left(  \varepsilon_{n_{1}}%
-\varepsilon_{n_{2}}\right)  }  .
\end{align}

\section{Continuum model of twisted bilayer MoTe$_2$}
To illustrate characteristics of the linear polarization effect, we first consider R-stacked twisted and strained bilayer MoTe$_2$ as an example.
Specifically, we apply an uniaxial strain along the zigzag direction of the top layer and twist the two layers by opposite angles. The purpose of strain is to break the rotational symmetries.
In the following, we use $\theta$ and $\epsilon$ to denote the twist angle and strain intensity, respectively. Also, $\mathcal{R}_{\frac{\theta}{2}}$ and $\mathcal{R}_{-\frac{\theta}{2}}$ denote the rotation matrix on the top and bottom layer, and $\mathcal{S} = \text{diag}\{\epsilon,-\nu \epsilon\}$ with Poisson ratio $\nu = 0.16$ is the strain tensor acting on the top layer.

A four-band continuum model around $\mathbf{K}_\xi$ valley can be established ($\xi = \pm$)~\cite{Bi2019,ZhaiPRM2020,wu2019topological}:
\begin{eqnarray}
    \mathcal{H}_{\xi}(\mathbf{k}, \mathbf{r})=
    \begin{pmatrix}
        \mathcal{H}_{b, \xi}(\mathbf{k})+V_b(\mathbf{r}) & T_{\xi}(\mathbf{r}) \\
T_{\xi}^{\dagger}(\mathbf{r}) & \mathcal{H}_{t, \xi}(\mathbf{k})+V_t(\mathbf{r})
    \end{pmatrix}.
\label{formula_Hamiltonian_TMD}
\end{eqnarray}
In diagonal elements, the monolayer Hamiltonian reads~\cite{Bi2019,ZhaiPRM2020}
\begin{eqnarray}
    \mathcal{H}_{l, \xi}(\mathbf{k})=\hbar v_F\left[\mathcal{M}_l^T\left(\mathbf{k}-\mathbf{D}_{l, \xi}\right) \cdot\left(\xi \tau_x, \tau_y\right)\right]+\frac{\Delta_g}{2}\left(\mathcal{I}_2+\tau_z\right),
\end{eqnarray}
where Pauli matrices $\tau_{x,y,z}$ spans the atomic orbital degree-of-freedom, $\mathcal{I}_2$ is the identity matrix, $\mathcal{M}_t = \mathcal{R}_{\frac{\theta}{2}} (\mathcal{I}_2 + \mathcal{S})$, $\mathcal{M}_b = \mathcal{R}_{-\frac{\theta}{2}}$, $v_F = 0.4 \times 10^6$m/s and $\Delta_g = 1.1$eV are Fermi velocity and band gap~\cite{wu2019topological}. $\mathbf{D}_{l,\xi}$ is the monolayer Dirac point, which is shifted by strain towards
\begin{eqnarray}
    \mathbf{D}_{l, \xi}=(\mathcal{M}_l^{-1})^T \mathbf{K}_{\xi}-\xi \mathbf{A}_S
\end{eqnarray}
where $\mathbf{A}_\text{S} = \frac{\sqrt{3}\beta}{2a}[(1+\nu)\epsilon,0]$ is the strain-induced vector potential, $a = 3.472$\AA \, is the monolayer lattice constant, and $\beta \approx 2.4$ is a material-dependent parameter~\cite{fang2018electronic}.
The second term in the diagonal elements is the energy modulation caused by the moir\'e landscape $V_l(\mathbf{r}) = \text{diag}\{V_{l,c}(\mathbf{r}),V_{l,v}(\mathbf{r})\}$, where $V_{l, c / v}(\mathbf{r})=2 V_{c / v}^0 \sum_{j=1,2,3} \cos \left(\mathbf{g}_j \cdot \mathbf{r}+s_l \varphi_{c / v}\right)$ describes the potential on conduction/valence bands, with $V^0_{c/v} = 8/5.97$meV, $\varphi_{c/v} = -87.9^\circ / -89.6^\circ$, and $s_{b/t} = \pm 1$~\cite{wu2019topological}.

The off-diagonal term in (\ref{formula_Hamiltonian_TMD}) represents the interlayer tunneling
\begin{eqnarray}
    T_{\xi}(\mathbf{r})=
    \begin{pmatrix}
        u_{c c} & u_{c v} \\
u_{v c} & u_{v v}
    \end{pmatrix}
+
\begin{pmatrix}
    u_{c c} & u_{c v} e^{i \xi 2 \pi / 3} \\
u_{v c} e^{-i \xi 2 \pi / 3} & u_{v v}
\end{pmatrix}
e^{i \xi \mathbf{g}_1 \cdot \mathbf{r}}+
\begin{pmatrix}
    u_{c c} & u_{c v} e^{-i \xi 2 \pi / 3} \\
u_{v c} e^{i \xi 2 \pi / 3} & u_{v v}
\end{pmatrix}
e^{-i \xi \mathbf{g}_2 \cdot \mathbf{r}},
\end{eqnarray}
where $u_{cc/vv} = -2 / -8.5$ meV, $u_{cv} = 15.3$ meV, and $u_{cv} = u^*_{vc}$~\cite{wu2019topological}. The moir\'e reciprocal lattice vectors $\mathbf{g}_{j=1,2}$ are constructed from the monolayer reciprocal lattice vectors $\mathbf{b}_j = \mathcal{R}_{\frac{2\pi}{3}}^{(j-1)}\frac{4\pi}{\sqrt{3}a}(\frac{\sqrt{3}}{2},\frac{1}{2})$ via $\mathbf{g}_j \approx (\mathcal{R}_\theta+\mathcal{S})^T\mathbf{b}_j$, and $\mathbf{g}_3 = -(\mathbf{g}_1+\mathbf{g}_2)$.

For calculating the electric polarization, we choose the interlayer distance to be $d=7$~\AA.


\section{Dipole responses at larger twist angles of a twisted bilayer}

\begin{figure*}[h]
	\includegraphics[width=1\textwidth]{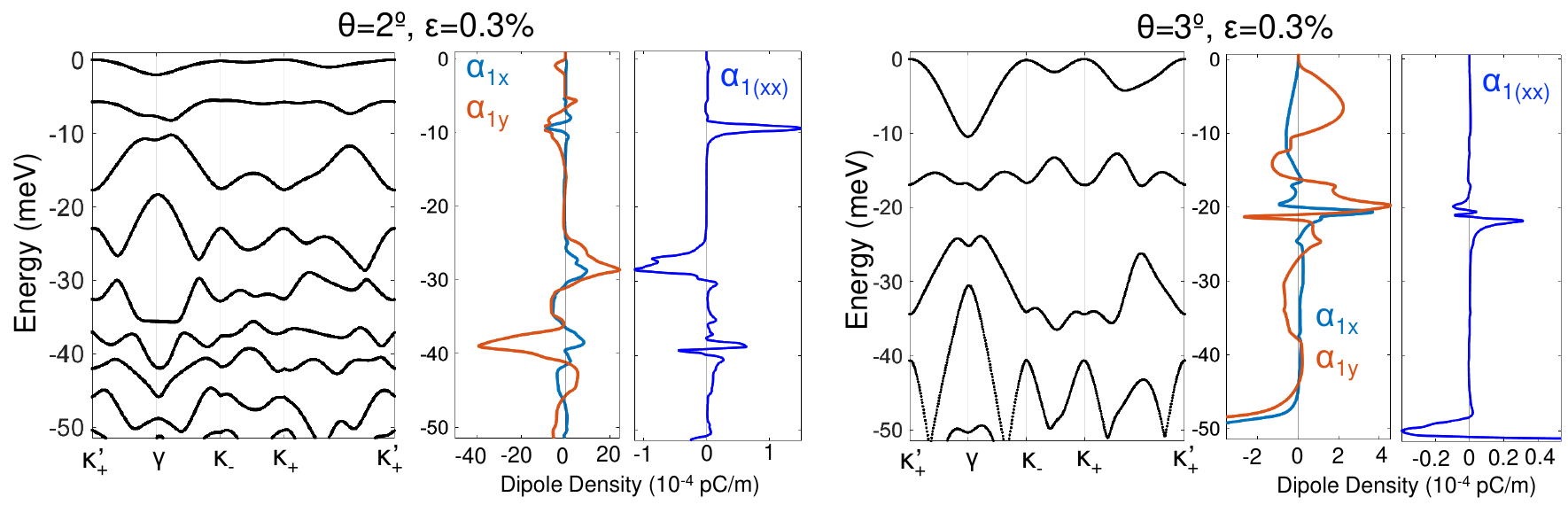} \caption{Linear and Nonlinear dipole response of MoTe$_2$ bilayer at twist angle $\theta = 2^\circ$ and $\theta = 3^\circ$, respectively. Strain intensity is fixed at $\epsilon = 0.3\%$.}
	\label{fig_largeangle}%
\end{figure*}

The multipole generation can exist in a very large range of twist angles. However, it is most significant when the electronic states are strongly layer-hybridized and the adjacent energy bands have small separations so that the geometric quantities are large. As the twist angle is increased, the moir\'e period decreases, and moir\'e Brillouin zone expands. This leads to band folding occurring at higher energies around which interlayer coupling effects are most pronounced, while the low-energy electronic states are less layer-hybridized accompanied by more dispersive bands. As a result, small twist angles are usually desirable for the multipole generation if one focuses on the low-energy regime. Fig.~\ref{fig_largeangle} compares the dipole generation in bilayer MoTe$_2$ at twist angle of $\theta = 2^\circ$ and $\theta = 3^\circ$. One can clearly identify that the energy bands become more dispersive with larger angles and the corresponding dipole density decreases in the low-energy region (e.g., from -20 to 0 meV). Despite the changes, the dipole densities in the two examples are still quite significant and can be experimentally measured.


\section{Continuum model of twisted trilayer MoTe$_2$ and layer polarization in the mirror-symmetric case\label{supp_trilayer}}
\subsection{Mirror-symmetric case}
\begin{figure*}
	\includegraphics[width=0.9\textwidth]{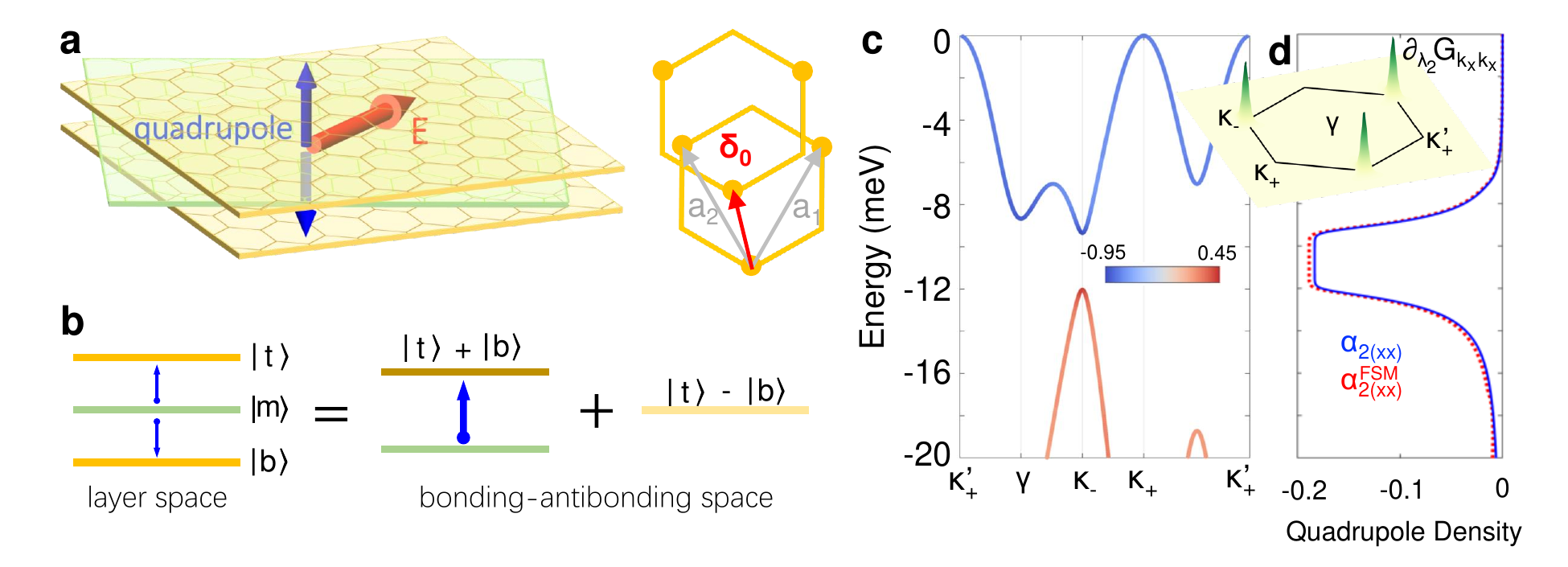} \caption{(a) Schematics of the quadrupole induced by in-plane $\bs{E}$ field on the twisted trilayer MoTe$_{2}$. The right panel illustrates the top layer displacement $\bs{\delta}_{0}$ with respect to the bottom layer. (b) Decomposition of the mirror-symmetric trilayer into an effective bilayer model in $\{ \ket{\text{t}} + \ket{\text{b}},\ket{\text{m}}\}$ space and a decoupled single layer model in $\{ \ket{\text{t}} - \ket{\text{b}}\}$ space, where the dipole in the effective bilayer is equivalent to a quadrupole in a trilayer. (c) Low-energy valence bands with $\theta=3^{\circ}$. The color denotes $\braket{\hat{\sigma}_z}$ in $\{ \ket{\text{t}} + \ket{\text{b}},\ket{\text{m}}\}$ space. (d) Quadrupole density (units: $10^{-13}$ pC) as a function of Fermi energy. Blue and red dashed curves represent the exact result and FSM contribution of the $xx$ component, respectively. The plateau value of the quadrupole density corresponds to a voltage difference around 0.2 mV between the top and middle layer. Inset: \textit{k}-space distribution of $\partial_{\lambda_2}G_{k_x k_x}$ on the first valence band with arbitrary unit. The electric field is taken as $10^{4}$ V/m.}
	\label{fig_symmtrilayer}%
\end{figure*}

We now investigate the layer polarization effect of a twisted trilayer MoTe$_2$. Starting from a fully aligned trilayer configuration, the top and bottom layers are twisted simultaneously by angle $\theta/2$, while the middle layer is twisted inversely (Fig.~\ref{fig_symmtrilayer}a).  The periodic local stacking registries will appear in the top-middle and bottom-middle interfaces, and two aligned moir\'e patterns will form.

Based on Refs.~\cite{wu2019topological,tong2020interferences}, a continuum model for the valence band of MoTe$_2$ in $\mathbf{K}_{\xi}$ valley can be established,
\begin{equation}
    \mathcal{H}^{\text{trilayer}}_\xi(\mathbf{k},\mathbf{r}) = 
    \begin{pmatrix}
        -\frac{\hbar^2 (\mathbf{k}-\mathbf{K}_{t,\xi})^2}{2 m_{\text{eff}}} & 0 & 0 \\
        0 & -\frac{\hbar^2 (\mathbf{k}-\mathbf{K}_{m,\xi})^2}{2 m_{\text{eff}}} & 0 \\
        0 & 0 & -\frac{\hbar^2 (\mathbf{k}-\mathbf{K}_{t,\xi})^2}{2 m_{\text{eff}}}
    \end{pmatrix} +
    \begin{pmatrix}
        0 & h^*_\xi(\mathbf{r}) & 0 \\
        h_\xi(\mathbf{r}) & -2\mathcal{V}(\mathbf{r}) & h_\xi(\mathbf{r}) \\
        0 & h^*_\xi(\mathbf{r}) & 0
    \end{pmatrix}.
    \label{formula_trilayerH}
\end{equation}
where $\mathbf{k}$ is the wavevector away from the Dirac points, $m_{\text{eff}} = 0.62 m_e$ is effective mass, $m_e$ the free electron mass. $\mathbf{K}_{t/m,\xi} = \mathcal{R}_{\pm \frac{\theta}{2}} \mathbf{K}_\xi$. $h_\xi$ and $\mathcal{V}$ are interlayer coupling and electrostatic potential, respectively:
\begin{align}
    h_\xi(\mathbf{r}) &= u_{vv}(1+e^{i \xi \mathbf{g}_1 \cdot \mathbf{r}} + e^{-i \xi \mathbf{g}_2 \cdot \mathbf{r}}), \\
    \mathcal{V}(\mathbf{r}) &= -2 V_v^0 \text{sin}(\varphi_v) \sum_{j=1}^3 \text{sin}(\mathbf{g}_j \cdot \mathbf{r}).
\end{align}

Mirror symmetry $\mathcal{M}_z$ of the structure forbids the dipole polarization, but it permits the formation of quadrupoles. 
Before looking at the numerical results, we point out that the nonlinear quadrupole generation in the presence of $\mathcal{M}_z$ symmetry can also be understood from the perspective of an effective nonlinear dipole response~\cite{TrilayerNanoLett2020}.
The above Hamiltonian~(\ref{formula_trilayerH}) is in the layer basis $\left\{ \ket{\text{t}},\ket{\text{m}},\ket{\text{b}} \right\}$, by transforming into the bonding-antibonding basis $\left\{ \ket{+},\ket{\text{m}},\ket{-} \right\}$ through
\begin{equation}
    \mathcal{O} = \frac{1}{\sqrt{2}}
    \begin{pmatrix}
        1 & 0 & 1 \\
        0 & \sqrt{2} & 0 \\
        1 & 0 & -1
    \end{pmatrix},
\end{equation}
the trilayer can be equivalently reduced to an effective bilayer in $\left\{ \ket{\text{t}} + \ket{\text{b}},\ket{\text{m}} \right\}$ space and a decoupled monolayer in $\left\{ \ket{\text{t}} - \ket{\text{b}} \right\}$ space (Fig.~\ref{fig_symmtrilayer}b),
\begin{equation}
\begin{aligned}
    \tilde{\mathcal{H}}^{\text{trilayer}}_\xi &= 
    \mathcal{O}^\dagger \mathcal{H}^{\text{trilayer}}_\xi \mathcal{O} \\
    & = 
    \begin{pmatrix}
        -\frac{\hbar^2 (\mathbf{k}-\mathbf{K}_{t,\xi})^2}{2 m_{\text{eff}}} & 0 & 0 \\
        0 & -\frac{\hbar^2 (\mathbf{k}-\mathbf{K}_{m,\xi})^2}{2 m_{\text{eff}}} & 0 \\
        0 & 0 & -\frac{\hbar^2 (\mathbf{k}-\mathbf{K}_{t,\xi})^2}{2 m_{\text{eff}}}
    \end{pmatrix} +
    \begin{pmatrix}
        0 & \sqrt{2}h^*_\xi(\mathbf{r}) & 0 \\
        \sqrt{2}h_\xi(\mathbf{r}) & -2\mathcal{V}(\mathbf{r}) & 0 \\
        0 & 0 & 0
    \end{pmatrix}.
\end{aligned}
\end{equation}
The effective bilayer model $\mathcal{H}^{\text{eff}}_\xi = \text{diag}\left \{-\frac{\hbar^2 (\mathbf{k}-\mathbf{K}_{t,\xi})^2}{2 m_{\text{eff}}},-\frac{\hbar^2 (\mathbf{k}-\mathbf{K}_{m,\xi})^2}{2 m_{\text{eff}}} \right \}+ 
\begin{pmatrix}
    0 & \sqrt{2} h^*_\tau \\
    \sqrt{2} h_\tau & -2\mathcal{V}
\end{pmatrix}$ has an enlarged interlayer coupling and an asymmetric electrostatic potential.
Therefore, we can investigate the layer Edelstein effect of a trilayer by calculating the dipole response on the effective bilayer, where the ultimate electric polarization will behave as an electric quadruple on the homotrilayer. Besides, the symmetry analysis dictates that the trilayer belongs to $C_{3h}$ group and the effective bilayer belongs to $C_3$ point group, which forbids the dipole generation and the linear quadrupole generation, only the nonlinear quadrupole coefficient is allowed, which exhibits an isotropic feature with $\alpha_{xx} = \alpha_{yy}$, $\alpha_{xy} = \alpha_{yx} = 0$.

It is also important to point out some subtle differences between the bilayer and trilayer cases. The moir\'e potential contains two parts: intralayer potential and interlayer coupling term. The former localizes electrons, while the latter causes interlayer hybridization. As shown above, the intralayer potential of the middle (top/bottom) layer in a trilayer is stronger (suppressed) compared to that in a bilayer. This leads to localization towards the middle layer for low-energy electrons. Therefore, small twist angles are less favorable for multipole generation in a trilayer, which would otherwise lead to highly localized electronic states with negligible interlayer hybridization. Consequently, intermediate twist angles (e.g., $\theta \sim 3^\circ$) are most desirable in the case of trilayers.

Figure~\ref{fig_symmtrilayer}c shows the low-energy band structures of the $3^\circ$ twisted MoTe$_{2}$ homotrilayer.
The color denotes $\braket{\hat{\sigma}_z}$ in $\{ \ket{\text{t}} + \ket{\text{b}},\ket{\text{m}}\}$ space.
The electrons from the first (second) valence band are dominantly located in the middle (outer) layer. A finite gap appears between the two bands,
inside which there is a sizeable plateau characterizing the nonlinear quadrupole generation contributed by the $k$-space BCP (Fig.~\ref{fig_symmtrilayer}d). The regions around small gaps are expected to constitute the hot spots of BCP. Indeed, we find these regions support prominent distributions of band geometric quantity, as is shown in the inset of
Fig.~\ref{fig_symmtrilayer}d for the distribution of $\partial_{\lambda_2}G_{k_x k_x}$ on the first band. Also, as these two bands are relatively far away from other bands, the result is predominantly contributed by the quantum metric (see blue vs red dashed curves of Fig.~\ref{fig_symmtrilayer}d). The value of the quadrupole density can be converted into other more intuitive quantities. By taking the interlayer distance $d=7$\AA\, and an electric field of $10^4$ V/m, the peak quadrupole density in Fig.~\ref{fig_symmtrilayer}d corresponds to about 0.3 electrons per $\mu \text{m}^2$ by dividing $ed^2$, which generates a voltage difference $\sim O(0.1)$ mV between the outer and middle layers.

\subsection{General case}

The $\mathcal{M}_z$ mirror symmetry can be broken by displacing the top layer relative to the bottom layer. The in-plane displacement between the top and middle
layer is thus shifted: $\boldsymbol{{\delta}}_{tm} = \theta
\hat{z} \times \mathbf{r} + \boldsymbol{{\delta}}_0$. The constant displacement $\boldsymbol{{\delta}}_0$ adds a set of
phases in the electrostatic potential and interlayer tunneling between the top
and middle layer:
\begin{align}
  V_{\text{int}} (\boldsymbol{{\delta}}_{tm}) & =  - 2 V_v^0 \text{sin}
  (\varphi_v)  \sum_{j = 1}^3 \text{sin} (\mathbf{b}_j \cdot
  \boldsymbol{\delta}_{tm})
  =  - 2 V_v^0 \text{sin} (\varphi_v)  \sum_{j =
  1}^3 \text{sin} (\mathbf{g}_j \cdot \mathbf{r} + \mathbf{b}_j \cdot
  \boldsymbol{\delta}_0), \\
  h_{\xi} (\boldsymbol{{\delta}}_{tm}) & =  u_{v v}  (1 + e^{- i \xi
  \mathbf{b}_2 \cdot \boldsymbol{\delta}_{tm}} + e^{i \xi \mathbf{b}_1
  \cdot \boldsymbol{\delta}_{tm}}) 
   =  u_{v v}  (1 + e^{- i \xi
  \mathbf{b}_2 \cdot \boldsymbol{{\delta}}_0} e^{- i \xi \mathbf{g}_2 \cdot
  \mathbf{r}} + e^{i \xi \mathbf{b}_1 \cdot \boldsymbol{{\delta}}_0} e^{i \xi
  \mathbf{g}_1 \cdot \mathbf{r}}). 
\end{align}
The moir\'e potential for the general trilayer geometry can be built from such terms as 
\begin{eqnarray}
  \begin{pmatrix}
        V_{\text{int}} (\boldsymbol{{\delta}}_{tm}) - V_{\text{int}}
        (\boldsymbol{{\delta}}_{bm}) & h^*_{\xi}
        (\boldsymbol{{\delta}}_{tm}) & 0\\
        h_{\xi} (\boldsymbol{{\delta}}_{tm}) & - V_{\text{int}}
        (\boldsymbol{{\delta}}_{tm}) - V_{\text{int}}
        (\boldsymbol{{\delta}}_{bm}) & h_{\xi}(\boldsymbol{{\delta}}_{bm})\\
        0 & h_{\xi}^* (\boldsymbol{{\delta}}_{bm}) & - V_{\text{int}}
        (\boldsymbol{{\delta}}_{tm}) + V_{\text{int}}
        (\boldsymbol{{\delta}}_{bm})  
  \end{pmatrix}.
\end{eqnarray}
The extra phases $\mathbf{b}_{1,2,3} \cdot \boldsymbol{\delta}_0$ break the mirror and threefold rotational symmetry. Eq.~(\ref{formula_trilayerH}) is reproduced when $\boldsymbol{\delta}_0=0$.

The moire potential on each layer reads explicitly
\begin{align}
  V^{\text{tri}}_t & =  - V^{\text{tri}}_b = V_{\text{int}}
  (\boldsymbol{{\delta}}_{tm}) - V_{\text{int}}
  (\boldsymbol{{\delta}}_{bm}) 
   =  - 4 V_v^0 \text{sin} (\varphi_v)  \sum_{j = 1}^3 \cos \left(
  \mathbf{g}_j \cdot \mathbf{r} + \frac{1}{2} \mathbf{b}_j \cdot
  \boldsymbol{{\delta}}_0 \right) \sin \left( \frac{1}{2} \mathbf{b}_j \cdot
  \boldsymbol{{\delta}}_0 \right) \\
  V^{\text{tri}}_m & =  - V_{\text{int}} (\boldsymbol{{\delta}}_{tm}) -
  V_{\text{int}} (\boldsymbol{{\delta}}_{bm})
   =  4 V_v^0 \text{sin} (\varphi_v)  \sum_{j = 1}^3 \sin \left(
  \mathbf{g}_j \cdot \mathbf{r} + \frac{1}{2} \mathbf{b}_j \cdot
  \boldsymbol{{\delta}}_0 \right) \cos \left( \frac{1}{2} \mathbf{b}_j \cdot
  \boldsymbol{{\delta}}_0 \right) 
\end{align}
Specifically,
\begin{align*}
     V^{\text{tri}}_t &\sim - \sin
(\mathbf{g}_1 \cdot \mathbf{r}) - \sin (\mathbf{g}_3 \cdot \mathbf{r}),
V^{\text{tri}}_m \sim \sin (\mathbf{g}_2 \cdot \mathbf{r}),~\text{when } \boldsymbol{\delta}_0= 0.5 \mathbf{a}_1\\
V^{\text{tri}}_t &\sim - \sin
(\mathbf{g}_2 \cdot \mathbf{r}) - \sin (\mathbf{g}_3 \cdot \mathbf{r}),
V^{\text{tri}}_m \sim \sin (\mathbf{g}_1 \cdot \mathbf{r}),~\text{when } \boldsymbol{\delta}_0= 0.5 \mathbf{a}_2\\
V^{\text{tri}}_t&\sim - \sin (\mathbf{g}_1 \cdot \mathbf{r}) - \sin (\mathbf{g}_2 \cdot
\mathbf{r}), V^{\text{tri}}_m \sim \sin (\mathbf{g}_3 \cdot \mathbf{r}),~\text{when }
\boldsymbol{{\delta}}_0 = 0.5 \mathbf{a}_1 + 0.5 \mathbf{a}_2
\end{align*}
In such cases, some of the energy bands become degenerate  (e.g., the first two valence bands in Fig.~3 in the main text). 

\newpage

\section{Comparison of the polarization response with the intrinsic layer polarization}
\begin{figure*}[h]
	\includegraphics[width=0.4\textwidth]{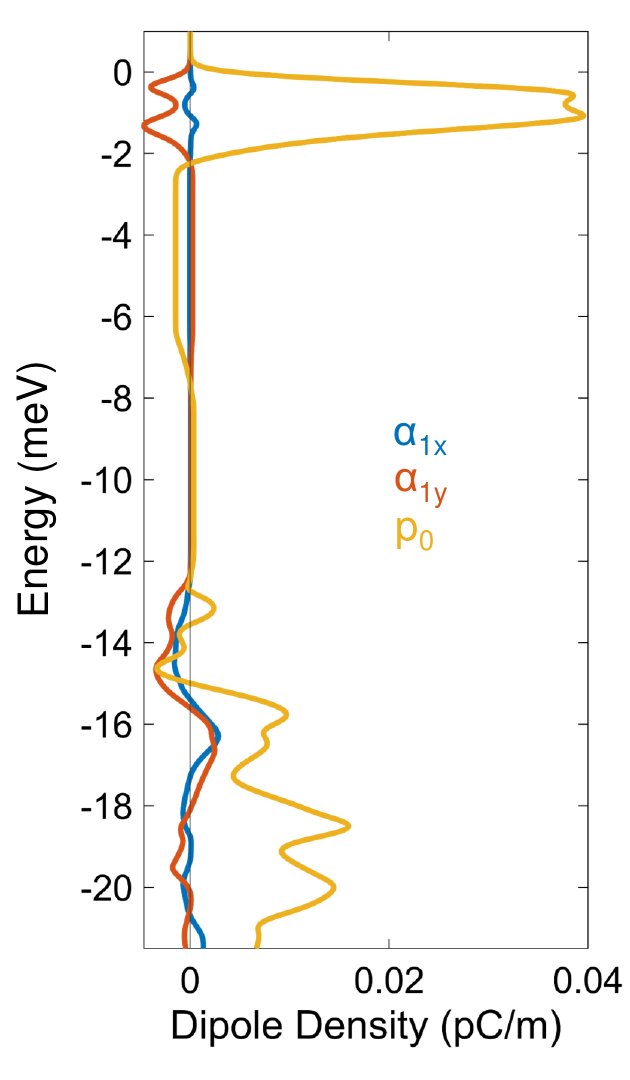} \caption{Intrinsic polarization as a function of chemical potential. The linear responses are the same as in Fig.~2c in the main text. The units of background polarization and linear responses have been unified.}
	\label{fig_p0}%
\end{figure*}

As a comparison, we show the intrinsic polarization of the $1.2^\circ$ MoTe$_2$ homobilayer with 0.3\% strain (Fig.~\ref{fig_p0}), the peak corresponds to $O(1)$~mV built-in voltage difference. For commensurate rhombohedral-stacked TMDs, the built-in voltage difference of the XM structure (or the MX structure with opposite electric polarization) can reach 50-70 mV~\cite{wang2022interfacial}. Consider the mixture of AA, XM and MX domains in a moir\'e supercell, the $O(1)$~mV of the intrinsic voltage difference from our calculation is reasonable.

\bibliography{LEE_ref}